\documentclass[11pt]{article}
\usepackage[T1]{fontenc}
\usepackage[utf8x]{inputenc}
\usepackage{geometry}
\usepackage{color}
\usepackage{booktabs}
\usepackage{amsmath}
\usepackage{amssymb}
\usepackage{graphicx}
\usepackage{setspace}
\usepackage[unicode=true,
 bookmarks=false,
 breaklinks=true,pdfborder={0 0 0},pdfborderstyle={},backref=false,colorlinks=true]
 {hyperref}
\hypersetup{
 urlcolor=blue,linkcolor=blue,citecolor=blue}

\makeatletter

\providecommand{\tabularnewline}{\\}

\usepackage{cite}

\makeatother

\begin{document}

\title{\vspace{0.5in}
Method of Manufactured Solutions Code Verification of Elastostatic
Solid Mechanics Problems in a Commercial Finite Element Solver}

\author{Kenneth I. Aycock{*}, Nuno Rebelo†, and Brent A. Craven{*}}
\maketitle
\begin{center}
{*}Division of Applied Mechanics, Office of Science and Engineering
Laboratories, Center for Devices and Radiological Health, United States
Food and Drug Administration, 10903 New Hampshire Avenue, Silver Spring,
MD 20993; †Nuno Rebelo Associates, LLC, Fremont, CA 94539
\par\end{center}
\begin{abstract}
Much progress has been made in advancing and standardizing verification,
validation, and uncertainty quantification practices for computational
modeling in recent decades. However, examples of rigorous code verification
for solid mechanics problems in the literature remain scarce, particularly
for commercial software and for the non-trivial large-deformation
analyses and nonlinear materials typically needed to simulate medical
devices. Here, we apply the method of manufactured solutions (MMS)
to verify a commercial finite element code for elastostatic solid
mechanics analyses using linear-elastic, hyperelastic (neo-Hookean),
and quasi-hyperelastic (Hencky) constitutive models. Analytical source
terms are generated using either \texttt{Python/SymPy} or \texttt{Mathematica}
and are implemented in \texttt{ABAQUS/Standard} without modification
to the solver source code. Source terms for the three constitutive
models are found to vary nearly six orders of magnitude in the number
of mathematical operations they contain. Refinement studies reveal
second-order displacement convergence in response to mesh refinement
for all constitutive models and first-order displacement convergence
in response to increment refinement for the finite-strain problems.
We also investigate the sensitivity of MMS convergence order to minor
coding errors using an exploratory case. Code used to generate the
MMS source terms and the input files for the simulations are provided
as supplemental material. \textbf{}
\end{abstract}

\section{Introduction}

Computational modeling and simulation (CM\&S) are anticipated to
play an increasingly significant role in the medical device industry
in the coming years \cite{morrison2018advancing}. Emerging applications
of CM\&S include: supporting claims of substantial equivalence or
safety and effectiveness of a medical device in regulatory submissions,
augmenting clinical trial data with evidence provided by ``virtual
clinical trials'' \cite{himes2016augmenting,himes2018use,haddad2014fracture},
and providing CM\&S–derived diagnostic information for clinical decision
support (e.g., \cite{taylor2013computational,zarins2013computed}).
Such uses of CM\&S should be accompanied with evidence of model credibility
that is commensurate with the risk associated with the intended context
of use (COU) of the model \cite{ASME_VV40}. Thus, as CM\&S extends
into higher-risk applications, there will be an increasing need for
demonstrating greater CM\&S credibility through rigorous verification,
validation, and uncertainty quantification (VVUQ) activities.

The first step in demonstrating the credibility of a physics-based
computational model is code verification, or the process of ensuring
that the software solves the underlying mathematical equations correctly
\cite{ASME-VV10,ASME_VV20,ASME_VV40,oberkampf2010verification,roache2009fundamentals}.
Rigorous methods for code verification include the method of manufactured
solutions (MMS), considered the gold-standard by experts in the field
\cite{roache2009fundamentals,oberkampf2010verification}, and the
method of exact solutions (MES).

The method of manufactured solutions is a mathematical approach used
to verify that a given code solves the underlying governing equations
correctly and that grid and time step (or pseudo-time increment) refinement
reduces numerical error at the expected rate based on the underlying
numerical schemes \cite{roache2002code}. The technique was first
used by Roache \cite{roache1972computational} and was later described
more formally by Steinberg and Roache \cite{steinberg1985symbolic}.
Oberkampf \cite{oberkampf1995methodology} also used the procedure
and coined the term ``method of manufactured solutions.'' Although
the majority of MMS applications to date have focused on verifying
codes for computational fluid dynamics (CFD), application of MMS to
solid mechanics software has also been performed to some extent, likely
beginning with Bathe et al. \cite{bathe1990displacement} who used
an ``ad-hoc problem” with prescribed displacements to verify the
order of accuracy for 2D plate finite elements. Batra and Liang \cite{batra1997finite}
and Batra and Love \cite{batra2006consideration} also used what they
called a ``method of fictitious body forces'' to verify finite element
codes, although only for error quantification and not for formal order
of accuracy checking. Solid mechanics MMS has received more attention
recently for finite deformation analyses (e.g., \cite{Chamberland_2010,kamojjala2015verification}),
albeit only for in-house codes where the users had direct access to
solver source code.

Here, we use MMS to verify a \emph{commercial} finite element code
for elastostatic solid mechanics problems. We also discuss the unique
considerations and challenges associated with performing MMS when
the source code is not available to the user \cite{salari2000code}.
Specifically, we consider three classes of problems:
\begin{itemize}
\item \textbf{Case I}: \emph{small strain, linear elasticity} 
\item \textbf{Case II}: \emph{finite strain, neo-Hookean hyperelasticity} 
\item \textbf{Case III}: \emph{finite strain, Hencky elasticity }\cite{xiao2002hencky,xiao2005hencky}
\end{itemize}
We follow a typical MMS workflow \cite{roache2002code,salari2000code}
that includes the following steps:
\begin{enumerate}
\item choose a relevant domain for performing verification (e.g., 2D or
3D, static or time-resolved) 
\item write the governing equations for the problem of interest 
\item choose an analytical solution for a field of interest in the governing
equations (note that the solution does not need to be physically meaningful
\cite{roache2002code,roache2009fundamentals}) 
\item generate an analytical source term (i.e., a fictitious body force)
to satisfy the governing equations from (2) with the analytical solution
from (3) 
\item implement the source term and the appropriate boundary conditions
in simulation input files
\item perform grid or time/increment convergence studies and quantify the
observed order of convergence ($\textrm{OOC}_{\textrm{obs}}$)
\item compare the observed ($\textrm{OOC}_{\textrm{obs}}$) and theoretical
($\textrm{OOC}_{\textrm{theor}}$) convergence rates of the underlying
numerical algorithm. 
\end{enumerate}
Steps 1-4 are performed using a symbolic computer algebra system (\texttt{Python/SymPy}
for Cases I and II and \texttt{Mathematica} for Case III). Source
terms are then incorporated into \texttt{ABAQUS} input files (step
5) and grid and pseudo-time increment refinement studies are performed
(step 6). Finally, the results from steps 6 and 7 are analyzed and
plotted using \texttt{Python}. We have made the code used to generate
the MMS source terms openly available online at \href{https://figshare.com/s/a67927162e674bbb791e}{https://figshare.com/s/a67927162e674bbb791e}
as a \texttt{Python Jupyter} notebook with the intent that others
may use this material as a starting point for performing MMS code
verification of their own solid mechanics software or applications
of interest.

\section{Methods}

\subsection{Choose the problem domain}

\begin{figure}
\begin{centering}
\includegraphics[width=0.95\textwidth,height=0.4\textheight,keepaspectratio]{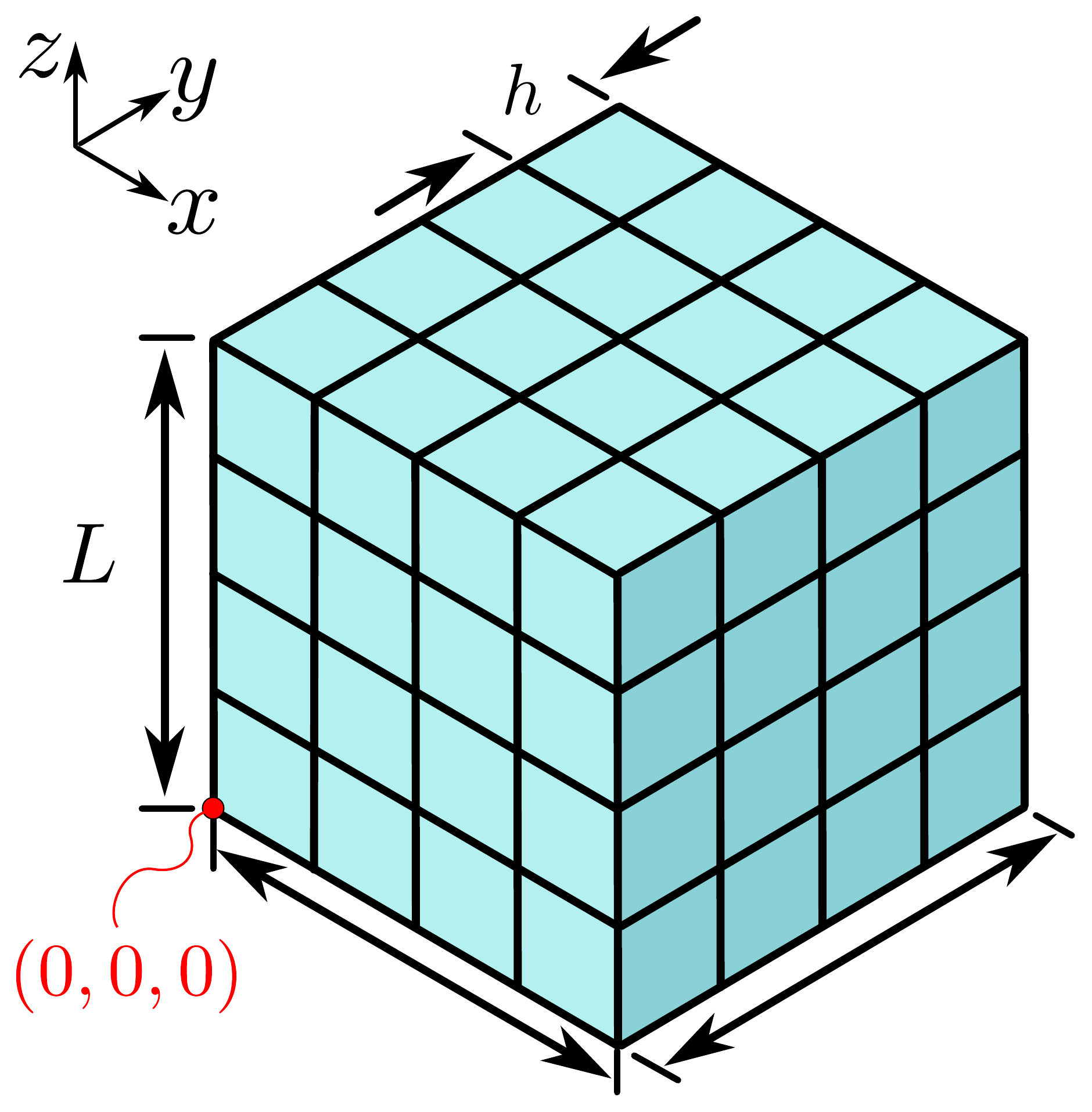}
\par\end{centering}
\caption{Unit cube domain with side lengths $L=1$ and grid spacing $h$.\label{fig:domain}}
\end{figure}

The MMS code verification is performed in a unit cube domain (reference
configuration $\mathbf{X}\in[0,1]^{3}$) with a grid spacing of $h$
(Fig. \ref{fig:domain}). All cases consider elastostatic deformation
($\frac{\partial}{\partial t}=0$).

\subsection{Write the governing equations}

The equations for conservation of linear momentum at static equilibrium,
written in the current configuration $\mathbf{x}$, are 
\begin{equation}
-\nabla_{\mathbf{x}}\cdot\boldsymbol{\sigma}\left(\mathbf{x}\right)=\mathbf{\mathbf{b}}\left(\mathbf{x}\right)\label{eq:govn-eulerian}
\end{equation}
where $\nabla_{\mathbf{x}}\cdot$ is the \emph{spatial} divergence
operator, $\boldsymbol{\sigma}$ is the Cauchy stress tensor, and
$\mathbf{b}$ is a body force vector with units of force per volume
(both in the current configuration). To facilitate the calculation
of the source term as a function of the reference configuration $\mathbf{X}$,
the conservation equations are pulled back to the reference configuration
using Nanson's relation \cite{Truesdell1960,bathe2006finite} and
written as
\begin{align}
-\nabla_{\mathbf{X}}\cdot\left(J\,\boldsymbol{\sigma\left(\mathbf{X}\right)}\,\mathbf{F}^{-T}\right) & =J\,\mathbf{b}\left(\mathbf{X}\right)\\
-\nabla_{\mathbf{X}}\cdot\mathbf{P}\left(\mathbf{X}\right) & =J\,\mathbf{b}\left(\mathbf{X}\right)\\
-\nabla_{\mathbf{X}}\cdot\mathbf{P}\left(\mathbf{X}\right) & =\boldsymbol{\phi}\left(\mathbf{X}\right)\label{eq:momentum-eqns}
\end{align}
where
\begin{equation}
\mathbf{P}\left(\mathbf{X}\right)=J\,\boldsymbol{\sigma}\left(\mathbf{X}\right)\,\mathbf{F}^{-T}\label{eq:PKI}
\end{equation}
is the first Piola-Kirchhoff stress tensor, which maps forces in the
current configuration to geometry in the reference configuration;
$\mathbf{F}$ is the deformation gradient; $J$ is the determinant
of $\mathbf{F}$, a measure of volume change; $\nabla_{\mathbf{X}}\cdot$
is the \emph{material} divergence operator; and $\boldsymbol{\phi}\left(\mathbf{X}\right)=J\:\mathbf{b}$
is a source term with units of force (current configuration) per volume
(\emph{reference} configuration). 

For problems using the small-strain approximation, \texttt{ABAQUS}
uses the simplification that the Cauchy stress tensor is equal to
the first Piola-Kirchhoff stress (see\emph{ Reference Library > Abaqus
> Theory > Elements > Continuum elements > Solid element formulation}
\cite{simulia2016abaqus}). Therefore, for the small-strain case (Case
I), 
\begin{equation}
\boldsymbol{\sigma}\left(\mathbf{X}\right)\approx\mathbf{P}\left(\mathbf{X}\right)\label{eq:small-strain-approximation}
\end{equation}
and 
\begin{equation}
-\nabla_{\mathbf{X}}\cdot\boldsymbol{\sigma}\left(\mathbf{X}\right)=\boldsymbol{\phi}\left(\mathbf{X}\right).
\end{equation}

To perform MMS, we will now choose a displacement field $\mathbf{u}\left(\mathbf{X}\right)$
and solve for the source term $\boldsymbol{\phi}\left(\mathbf{X}\right)$,
which is conveniently calculated based on the reference configuration
alone.

\subsection{Choose an analytical solution for the displacement}

\begin{figure}
\begin{centering}
\includegraphics[width=0.95\textwidth,height=0.4\textheight,keepaspectratio]{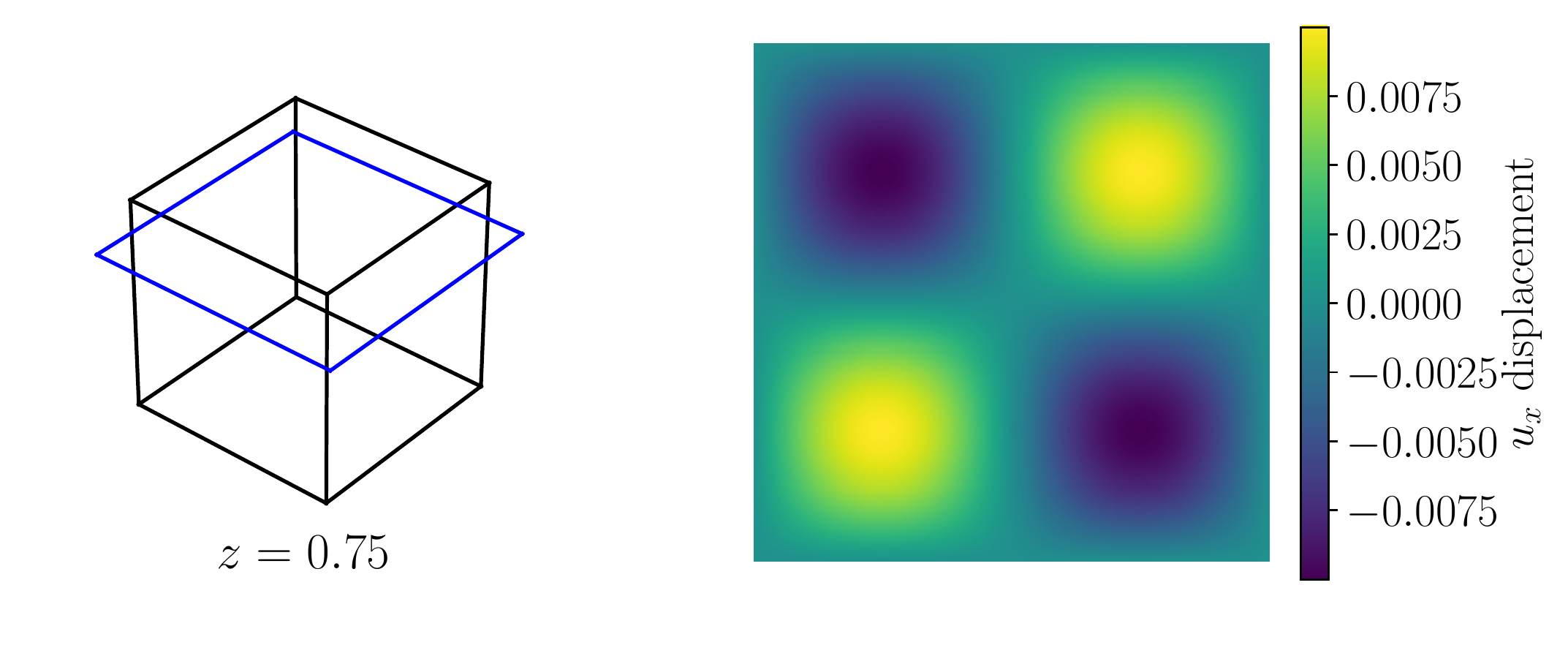}
\par\end{centering}
\caption{Contour plot showing $u_{x}$ displacement on the $z=3/4$ plane with
$C_{1}=0.01$ and $n=2$.\label{fig:example-disp}}
\end{figure}

The displacement vector $\mathbf{u}$ describes the mapping between
the current and reference configurations via the relationship $\mathbf{x}=\mathbf{X}+\mathbf{u}$.
Following the MMS described by Elsworth \cite{elsworth2014verification},
we choose the infinitely differentiable displacement field 
\begin{equation}
\mathbf{u}=C_{1}\sin\left(n\pi X\right)\sin\left(n\pi Y\right)\sin\left(n\pi Z\right)\left(\begin{array}{c}
1\\
1\\
1
\end{array}\right)\label{eq:u-MMS}
\end{equation}
where $C_{1}$ controls the magnitude of the displacement and $n$
controls the number of periods within the domain. Note that the displacement
at the boundaries is (conveniently) zero in all directions (e.g.,
Fig. \ref{fig:example-disp}). Accordingly, a fixed zero-displacement
boundary condition is enforced on the boundaries of the domain when
performing simulations.

\subsection{Generate the analytical source term to satisfy the governing equations}

\subsubsection{Common equations}

Solution of Eqn. \ref{eq:govn-eulerian} requires that the Cauchy
stress is expressly written as some function of material deformation,
known as the constitutive relation or model. Most constitutive models
make use of the deformation gradient tensor 
\begin{align}
\mathbf{F} & =\frac{\partial\mathbf{x}}{\partial\mathbf{X}}\\
 & =\mathbf{I}+\nabla_{\mathbf{X}}\mathbf{u}
\end{align}
where $\mathbf{I}$ is the second order identity tensor and $\nabla_{\mathbf{X}}\mathbf{u}$
is the material displacement gradient tensor. Many constitutive models
also relate stress to the volume change using the quantity 
\begin{equation}
J=\det\mathbf{F}\,.
\end{equation}

Constitutive models for finite-strain problems require frame indifference
so that arbitrary rotations do not produce erroneous stresses \cite{gurtin1982introduction}.
However, the deformation gradient $\mathbf{F}$ is not frame-indifferent
and is not suitable for analyses involving large deformations or rotations.
For these problems, a frame-indifferent tensor such as the right or
left Cauchy-Green deformation tensors,
\begin{equation}
\mathbf{C}=\mathbf{F}^{T}\mathbf{F}
\end{equation}
and
\begin{equation}
\mathbf{B}=\mathbf{F}\mathbf{F}^{T}
\end{equation}
respectively, can be used. These tensors appropriately quantify normal
and shear deformations while remaining unaffected by rigid body rotations.

\subsubsection{Case I: small-strain linear elasticity}

For problems with infinitesimal strains and rotations, the small-strain
tensor 
\begin{align}
\boldsymbol{\epsilon} & =\frac{1}{2}\left(\mathbf{F}+\mathbf{F}^{T}\right)-\mathbf{I}\\
 & =\frac{1}{2}\left(\nabla_{\mathbf{X}}\mathbf{u}+\nabla_{\mathbf{X}}\mathbf{u}^{T}\right)
\end{align}
is used to quantify material deformation. The constitutive model for
a homogeneous linear-elastic material takes on the generalized form
of Hooke's law,
\begin{equation}
\boldsymbol{\sigma}=2\,\mu\,\boldsymbol{\epsilon}+\lambda\,\textrm{tr}\,\boldsymbol{\epsilon}\,\mathbf{I}\,,
\end{equation}
where $\lambda$ and $\mu$ are the first and second Lamé parameters,
respectively. Note that given $\lambda$ and $\mu$, the Young's modulus
$E$ and Poisson's ratio $\nu$ required by \texttt{ABAQUS} can be
calculated as $E=\frac{\mu\left(3\lambda+2\mu\right)}{\lambda+\mu}$
and $\nu=\frac{\lambda}{2\left(\lambda+\mu\right)}$.

\subsubsection{Case II: finite-strain, neo-Hookean hyperelasticity}

For the isothermal, finite-strain neo-Hookean model in \texttt{ABAQUS},
the strain energy density function is defined as \cite{simulia2016abaqus}
\begin{equation}
W=C_{10}\left(J^{-\frac{2}{3}}\,I_{1}-3\right)+\frac{1}{D_{1}}\left(J-1\right)^{2}\label{eq:W}
\end{equation}
where $C_{10}$ and $D_{1}$ are material constants and $I_{1}=\textrm{tr}\,\mathbf{B}$
is the first invariant of $\mathbf{B}$. Eqn. \ref{eq:W} is used
to calculate the second Piola-Kirchhoff stress tensor
\begin{equation}
\mathbf{S}=2\frac{\partial W}{\partial\mathbf{C}}\label{eq:PKII}
\end{equation}
which is then used to calculate the Cauchy stress via the relationship
\begin{equation}
\mathbf{\boldsymbol{\sigma}}=J^{-1}\mathbf{F}\mathbf{S}\mathbf{F}^{T}\,.\label{eq:sigma-vs-PKII}
\end{equation}
Combining Eqns. \ref{eq:W}–\ref{eq:PKII} and simplifying the expression
obtains
\begin{equation}
\mathbf{\boldsymbol{\sigma}}=2\,C_{10}\,J^{-5/3}\left(\mathbf{B}-\frac{1}{3}\,I_{1}\,\mathbf{I}\right)+\frac{2}{D_{1}}\left(J-1\right)\mathbf{I}\,.\label{eq:sigma-neo-1}
\end{equation}

The material constants used in \texttt{ABAQUS} for the neo-Hookean
model are related to the first and second Lamé parameters of a material
linearized at the origin as follows: $C_{10}=\frac{\mu}{2}$, $K_{0}=\lambda+\frac{2\mu}{3}$,
and $D_{1}=\frac{2}{K_{0}}$ \cite{simulia2016abaqus}. Using these
relationships Eqn. \ref{eq:sigma-neo-1} becomes 
\begin{equation}
\boldsymbol{\sigma}=\mu\,J^{-5/3}\left(\mathbf{B}-\frac{1}{3}\,I_{1}\,\mathbf{I}\right)+\frac{3\,\lambda+2\,\mu}{3}\left(J-1\right)\mathbf{I}
\end{equation}
providing the Cauchy stress as a function of displacement-based quantites
and Lamé parameters alone (i.e., $\boldsymbol{\sigma}=f\left(\mathbf{u},\,\lambda,\,\mu\right)$
since $\left\{ \mathbf{B},\:J,\,I_{1}\right\} =f\left(\mathbf{u}\right)$).

\subsubsection{Case III: finite-strain, Hencky elasticity}

For finite-strain Hencky elasticity, the logarithmic or Hencky strain
tensor 
\begin{equation}
\boldsymbol{\epsilon}_{\ln}=\ln\,\mathbf{V}
\end{equation}
is used to quantify material deformation, where 
\begin{equation}
\mathbf{V}=\sqrt{\mathbf{B}}
\end{equation}
is the left Cauchy stretch tensor. Note that here $\sqrt{\,\,}$ and
$\ln$ are matrix operations (rather than element-wise operations),
where $\ln$ is the principal matrix logarithm. Use of $\mathbf{V}$
in $\boldsymbol{\epsilon}_{\ln}$ makes the strain measure invariant
to rigid body rotations and, thus, suitable for analyses with large
deformation and rotation.

For the full 3D, finite-strain, Hencky-elastic problem, \texttt{Python/SymPy}
alone cannot generate the source term, as \texttt{SymPy} fails to
diagonalize the left Cauchy deformation tensor. Instead, \texttt{Mathematica}
(v11.0, Wolfram Research, Inc., Champaign, IL) is used to directly
perform the principal matrix logarithm symbolically. Because the resulting
expression is already quite large at this intermediate stage (leaf
count \cite{wolfram2011mathematica} $\approx$ 500,000), we also
perform the remainder of the symbolic calculation for this case in
\texttt{Mathematica}.

The constitutive model for finite-strain Hencky elasticity is the
same as that for the small-strain case, with the exception that $\boldsymbol{\epsilon}$
is replaced by $\boldsymbol{\epsilon}_{\ln}$, yielding 
\begin{equation}
\boldsymbol{\sigma}=2\,\mu\,\boldsymbol{\epsilon}_{\ln}+\lambda\,\textrm{tr}\,\boldsymbol{\epsilon}_{\ln}\,\mathbf{I}\,\textrm{,}
\end{equation}
an early form of Hencky elasticity that is quasi-hyperelastic \cite{xiao2002hencky,xiao2005hencky,bavzant2012work}.

\subsubsection{Equation summary}

The constitutive model relating the Cauchy stress tensor to material
deformation is defined as one of the following, 

\[
\boldsymbol{\sigma}=\begin{cases}
\begin{aligned} & 2\,\mu\,\boldsymbol{\epsilon}+\lambda\,\textrm{tr}\,\boldsymbol{\epsilon}\,\mathbf{I} &  & \text{Case I}\\
 & \mu\,J^{-5/3}\left(\mathbf{B}-\frac{1}{3}\,I_{1}\,\mathbf{I}\right)+\frac{3\,\lambda+2\,\mu}{3}\left(J-1\right)\mathbf{I} &  & \text{Case II}\\
 & 2\,\mu\,\boldsymbol{\epsilon}_{\ln}+\lambda\,\textrm{tr}\,\boldsymbol{\epsilon}_{\ln}\,\mathbf{I} &  & \text{Case III\,.}
\end{aligned}
\end{cases}
\]
The first Piola-Kirchhoff stress tensor is then 

\begin{equation}
\mathbf{P}=\begin{cases}
\boldsymbol{\sigma} & \text{Case I}\\
J\,\boldsymbol{\sigma}\,\mathbf{F}^{-T} & \text{Cases II \& III}\,.
\end{cases}\label{eqn:p-cases}
\end{equation}

\subsubsection{Source term calculation}

The first Piola-Kirchhoff stress tensor $\mathbf{P}$ from Eqn. \ref{eqn:p-cases}
is used to calculate $\boldsymbol{\phi}$, the fictitious body force
(i.e., the source term) needed to satisfy the momentum equations (Eqn.
\ref{eq:momentum-eqns}). The source term is calculated as 
\begin{equation}
\boldsymbol{\phi}=-\nabla_{\mathbf{X}}\cdot\mathbf{P}\left(\mathbf{X}\right)
\end{equation}
where $\nabla_{\mathbf{X}}\cdot$ is the material divergence.

To calculate the source term expression, constants (Table\textbf{
\ref{table:constants}}) are substituted for the displacement field
and the constitutive model, and the source term expression is simplified,
yielding 
\begin{equation}
\boldsymbol{\phi}=\boldsymbol{f}\left(\mathbf{X}\right)\,.
\end{equation}
That is, we obtain $\boldsymbol{\phi}$ as a function of the \emph{initial}
reference position $\mathbf{X}$ alone. With the chosen source term
constants (Table\textbf{ \ref{table:constants}}), two periods are
contained within the unit cube domain, and peak principal strains
are approximately $10\%$. 
\begin{table}
\caption{Material and source term constants used in the MMS cases.\label{table:constants}}
\vspace{5mm}

\centering{}%
\begin{tabular}{cc|cc|cc}
\hline 
$C_{1}$  & $n$  & $\lambda$  & $\mu$  & $E$  & $\nu$ \tabularnewline
\hline 
0.01 & 2  & 100  & 50  & $400/3$ & $1/3$ \tabularnewline
\hline 
\end{tabular}
\end{table}

\subsection{Implement the source term in \texttt{ABAQUS}}

We implement the source term in \texttt{ABAQUS} by prescribing loads
in the reference configuration $\mathbf{X}$ using both distributed
body forces (\texttt{{*}DLOAD}) and concentrated point loads (\texttt{{*}CLOAD}).
The \texttt{{*}DLOAD} subroutine uses element shape functions to distribute
forces over element volumes, whereas \texttt{{*}CLOAD} provides a
simpler interface for prescribing discrete forces directly onto nodes
of the finite element mesh. Although \texttt{{*}DLOAD} is the more
obvious choice for implementing the source term, there are subtle
nuances in the implementation of this subroutine in the small- and
finite-strain formulation that must be accounted for. Indeed, regardless
of the loading choice, correct implementation of the source term is
critical for useful MMS code verification.

\texttt{{*}DLOAD}: Note that the source term $\boldsymbol{\phi}$
has units of force (current configuration) per volume (\emph{reference}
configuration). \texttt{{*}DLOAD} uses compatible units of force per
volume, but the volume is that of the reference configuration when
\texttt{NLGEOM=OFF} and that of the \emph{current} configuration when
\texttt{NLGEOM=ON}. Therefore, when \texttt{NLGEOM=ON}, we must divide
$\boldsymbol{\phi}$ by $J$ (see Eqns. \ref{eq:govn-eulerian}–\ref{eq:momentum-eqns}),
i.e.
\begin{equation}
\textrm{DLOAD}=\begin{cases}
\phi & \textrm{NLGEOM=OFF}\\
\phi/J & \textrm{NLGEOM=ON}
\end{cases}\,.\label{eq:dload-and-nlgeom}
\end{equation}

\texttt{{*}CLOAD}: In contrast, the units of \texttt{{*}CLOAD} are
force. To spread the source term over the nodes of the mesh, we can
integrate $\boldsymbol{\phi}$ over the effective volume of each node,
$V_{n}$, or use the simplified discrete approximation $\boldsymbol{\phi}\,V_{n}$
evaluated at each node position in $\mathbf{X}$. For interior nodes
on a uniform mesh constructed from eight-node elements, the nodal
volumes in the reference configuration are conveniently equal to the
element volumes, as each interior node is shared by eight elements.
Thus, $V_{n}=h^{3}$, where $h$ is the grid spacing. The nodal volumes
for nodes on the exterior boundaries of the domain are different,
but the forces at the exterior nodes are irrelevant due to the fixed
zero-displacement boundary condition imposed on the outer boundaries.
Therefore, the appropriate point-load forces for implementing the
source term $\boldsymbol{\phi}$ are
\begin{equation}
\textrm{CLOAD}=\phi\,h^{3}\,.
\end{equation}
This simplistic discretization of the source term introduces somewhat
large numerical errors for the coarse starting meshes, but these errors
are reduced with mesh refinement such that error norms still converge
at a rate consistent with the underlying numerical method.

For Cases I and II, the source term is implemented in \texttt{ABAQUS}
input files using the \texttt{analytical field} option in \texttt{ABAQUS/CAE},
which accepts symbolic \texttt{Python} expressions directly. However,
the symbolic source term for Case III is extremely large (final leaf
count in excess of $10^{7}$), making the use of \texttt{CAE} impractical.
Thus for Case III the symbolic expression from \texttt{Mathematica}
is exported into a \texttt{Python}-compatible format (\texttt{Fortran}
syntax) and subsequently converted to \texttt{C} code using the \texttt{SymPy
autowrap} module \texttt{(unfuncify}). \texttt{ABAQUS} input lines
for either \texttt{{*}CLOAD} or \texttt{{*}DLOAD} are then generated
and written directly from \texttt{Python}.

\subsection{Perform grid convergence study and calculate observed order of convergence}

For a given numerical method, the total numerical error $\mathcal{E}_{\text{total}}$
is the sum of discretization errors $\mathcal{E}_{\text{disc}}$,
iterative convergence error $\mathcal{E}_{\text{iter}}$, and round-off
error $\mathcal{E}_{\text{round-off}}$,
\begin{equation}
\mathcal{E}_{\text{total}}=\mathcal{E}_{\text{spatial disc}}+\mathcal{E}_{\text{temporal disc}}+\mathcal{E}_{\text{iter}}+\mathcal{E}_{\text{round-off}}\,\textrm{.}
\end{equation}
Although simulations performed herein use the static form of the governing
equations, nonlinear finite-strain (\texttt{NLGEOM=ON}) analyses in
\texttt{ABAQUS} use a pseudo-time $t\in[0,\,1]$ over which the prescribed
boundary conditions are ramped. That is, \texttt{ABAQUS} does not
solve for the final equilibrium state directly but instead uses an
updated Lagrangian approach and solves the governing equations incrementally,
performing Newton–Raphson iterations to obtain a converged solution
at each pseudo-time $t$. Because this incremental approach generates
some numerical error, we expect the increment size to influence the
accuracy of the results for the finite-strain cases (Cases II and
III).

In a typical MMS analysis used to evaluate the spatial order of accuracy,
the code is verified by observing the order of convergence of the
total numerical error as the mesh is refined. This assumes that error
sources other than spatial discretization are negligible so that
\begin{equation}
\mathcal{E}_{\text{total}}\approx\mathcal{E}_{\text{spatial disc}}\,\textrm{.}\label{eq:numerical-error-spatial-MMS}
\end{equation}
For Eqn. \ref{eq:numerical-error-spatial-MMS} to hold, we must first
confirm that the simulations are relatively insensitive to increment
size (pseudo-time discretization), solver convergence tolerances,
and round-off error. To do so, we perform simulations in \texttt{ABAQUS/Standard}
(R2016x, Dassault Systèmes, Providence, RI) using \texttt{C3D8I} elements
and a grid spacing of $h=0.25$, standard solver tolerances, and a
fixed increment size. Solver tolerances and increment size are then
each independently reduced by an order of magnitude to investigate
their influences on the solution. We found that decreasing solver
tolerances yields an undetectable reduction in error norms while refining
the increment size from $dt=0.1$ to $dt=0.01$ yields error norm
reduction on the order of $0.1\%$. However, because the error norms
are smaller for the\texttt{ {*}DLOAD} cases, we found that the smaller
time increment size of $dt=0.01$ is required to maintain $\mathcal{E}_{\text{temporal disc}}\ll\mathcal{E}_{\text{spatial disc}}$
at the finest mesh level. In short, results from these preliminary
simulations confirm that Eqn. \ref{eq:numerical-error-spatial-MMS}
holds using the selected values and that the total numerical error
is approximately equal to the spatial discretization error.

\begin{figure}
\begin{centering}
\includegraphics[width=0.95\textwidth,height=0.95\textheight,keepaspectratio]{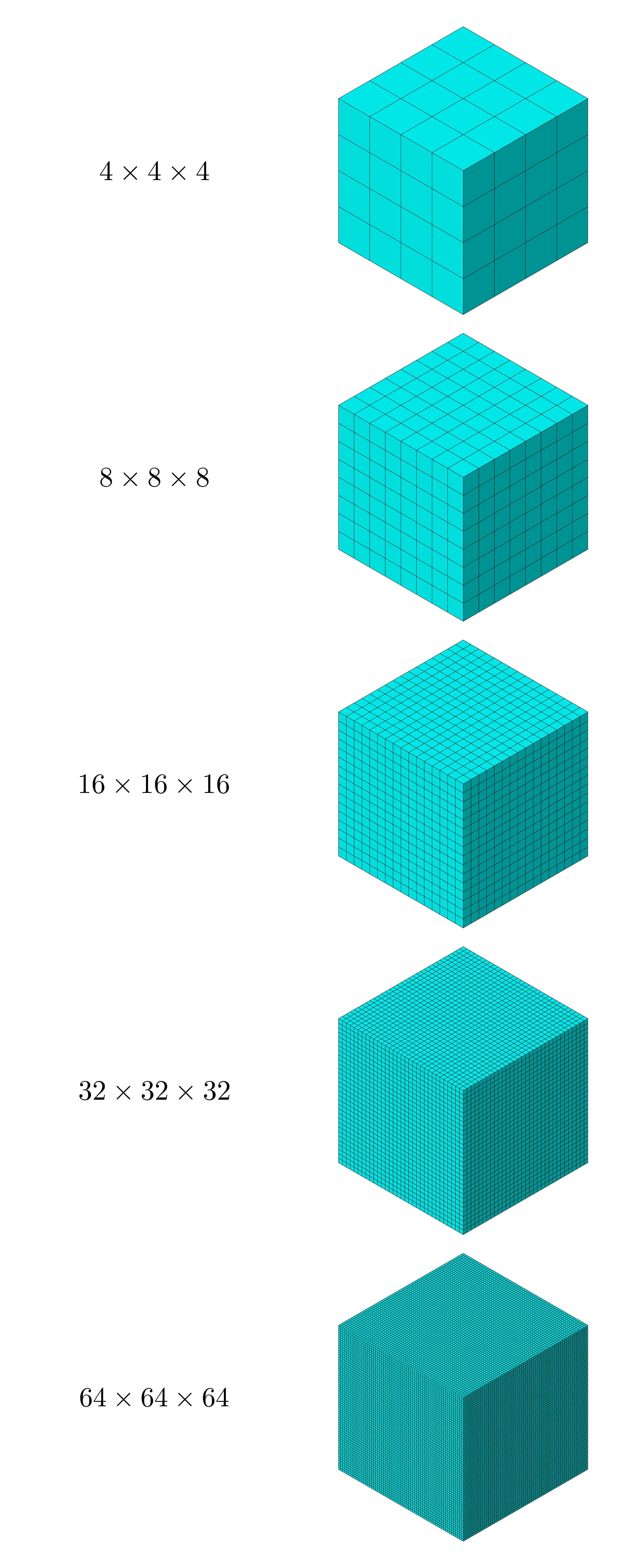}
\par\end{centering}
\caption{Meshes used in the grid refinement study.\label{fig:meshes} }
\end{figure}

Grid refinement studies are then performed as shown in Table \ref{table:simulation_summary}
for the three elastostatic cases using three common element types
(\texttt{C3D8}, \texttt{C3D8R}, and \texttt{C3D8I}) and two approaches
to implementing the source term (\texttt{{*}DLOAD} and \texttt{{*}CLOAD}).
Again, all grid refinement simulations are performed with standard
solver tolerances and a fixed increment size of $dt=0.1$ (\texttt{{*}CLOAD})
or $dt=0.01$ (\texttt{{*}DLOAD}). Uniform grid refinement is performed
using a total of five meshes (Fig. \ref{fig:meshes}). A consistent
refinement ratio of $r=2$ is used, where $r$ is the ratio of the
grid spacing from a one mesh to the next finer mesh (i.e., $r=\frac{h_{\textrm{coarser}}}{h_{\textrm{finer}}}$).
\begin{table}
\centering{}\caption{Summary of simulation settings for each case.\label{table:simulation_summary}}
\vspace{5mm}
\begin{tabular}{cccc}
\hline 
 & Case I  & Case II  & Case III\tabularnewline
\hline 
constitutive model  & linear elastic  & neo-Hookean  & Hencky elastic\tabularnewline
solver  & \texttt{Standard} & \texttt{Standard} & \texttt{Standard}\tabularnewline
\texttt{NLGEOM\textsuperscript{\texttt{1}}} & \texttt{OFF}  & \texttt{ON} & \texttt{ON}\tabularnewline
element types\textsuperscript{2} & \texttt{C3D8} & \texttt{C3D8} & \texttt{C3D8}\tabularnewline
 & \texttt{C3D8R} & \texttt{C3D8R} & \texttt{C3D8R}\tabularnewline
 & \texttt{C3D8I} & \texttt{C3D8I} & \texttt{C3D8I}\tabularnewline
load type\textsuperscript{2} & \texttt{{*}DLOAD} & \texttt{{*}DLOAD} & \texttt{{*}DLOAD}\tabularnewline
 & \texttt{{*}CLOAD} & \texttt{{*}CLOAD} & \texttt{{*}CLOAD}\tabularnewline
\hline 
\multicolumn{4}{l}{{\footnotesize{}\textsuperscript{1}In }\texttt{\footnotesize{}ABAQUS}{\footnotesize{},
the }\texttt{\footnotesize{}NLGEOM}{\footnotesize{} or ``nonlinear
geometry'' flag is used to activate the finite-strain formulation.}}\tabularnewline
\multicolumn{4}{l}{{\footnotesize{}\textsuperscript{2}Three element types and two source
term implementations are investigated independently.}}\tabularnewline
\end{tabular}
\end{table}
 
After performing simulations with each mesh resolution, the results
are extracted and displacement error norms are calculated. The normalized
Euclidean error magnitude 
\begin{equation}
\mathcal{E}_{\text{mag}}=\frac{1}{\sqrt{3}\,C_{1}}\sqrt{\sum_{i=1}^{3}\left(u_{\text{num}_{i}}-u_{\text{MMS}_{i}}\right)^{2}}
\end{equation}
is first calculated for all nodes, where $\sqrt{3}\,C_{1}$ is the
maximum displacement magnitude in the prescribed displacement field
(Eqn. \ref{eq:u-MMS}); $\boldsymbol{u}_{\text{num}}$ and $\boldsymbol{u}_{\text{MMS}}$
are the numerical and analytical displacement field vectors, respectively;
and $i$ is the $i^{\text{th}}$ component of the displacement vector.
The spatially averaged $L_{2}$ norm and the $L_{\infty}$ norm are
then calculated as 
\begin{equation}
L_{2}=\sqrt{\frac{1}{n}\sum_{i=1}^{n}\mathcal{E}_{\text{mag}}^{2}}
\end{equation}
and 
\begin{equation}
L_{\infty}=\max\left\{ \mathcal{E}_{\text{mag}}\right\} \,,
\end{equation}
where $n$ is the number of nodes in the domain. Finally, the observed
order of convergence ($\textrm{OOC}_{\textrm{obs}}$) of the total
numerical error is calculated as 
\begin{equation}
\textrm{OOC}_{\textrm{obs}}=\frac{\ln\left(L_{\text{c}}/L_{\text{f}}\right)}{\ln\left(r\right)}
\end{equation}
for each successive grid pair using both the $L_{2}$ or the $L_{\infty}$
error norms, where $L_{c}$ is a coarser mesh and $L_{f}$ a finer
mesh in a grid pair and $r$ is the refinement ratio. 

We also quantify the $\textrm{OOC}_{\textrm{obs}}$ of the pseudo-time
integration using a slightly different approach. To investigate and
quantify the importance of increment size, a separate increment sensitivity
study is performed using increment sizes of $dt=0.2$, $0.1$, $0.05$,
and $0.025$. A fixed grid spacing of $h=0.25$ is used for all cases
to hold the spatial discretization error constant. We then again quantify
the $L_{2}$ and $L_{\infty}$ error norms, but this time calculate
the rate of convergence of the error to an \emph{asymptotic but finite
value} since we cannot practically achieve $\mathcal{E}_{\text{spatial disc}}\ll\mathcal{E}_{\text{temporal disc}}$.
The convergence rate \cite{roache2009fundamentals} in response to
increment refinement is calculated as 
\begin{equation}
p=\frac{\ln\left(\frac{L_{\text{inc, c}}-L_{\text{inc, m}}}{L_{\textrm{inc, m}}-L_{\text{inc, f}}}\right)}{\ln\left(r_{\textrm{inc}}\right)}\,\textrm{,}
\end{equation}
where $L_{\textrm{inc}}$ is the error norm for a coarse (c), medium
(m), or fine (f) increment size. Note that $p$ is a function of the
\emph{relative} change\emph{ }in the total numerical error. Thus,
using a fixed mesh with a constant spatial discretization error, $p$
represents the rate at which the error due to increment discretization
alone is reduced to zero.

\section{Results}

\subsection{Source terms}

For Case I, small-strain linear elasticity, the source term $\boldsymbol{\phi}$
is relatively simple and can be expressed as
\begin{align}
\phi_{x} & =6\pi^{2}\left(2\sin{\left(2\pi X\right)}\sin{\left(2\pi Y\right)}\sin{\left(2\pi Z\right)}-\sin{\left(\pi\left(2Y+2Z\right)\right)}\cos{\left(2\pi X\right)}\right)\\
\phi_{y} & =6\pi^{2}\left(2\sin{\left(2\pi X\right)}\sin{\left(2\pi Y\right)}\sin{\left(2\pi Z\right)}-\sin{\left(\pi\left(2X+2Z\right)\right)}\cos{\left(2\pi Y\right)}\right)\\
\phi_{z} & =6\pi^{2}\left(2\sin{\left(2\pi X\right)}\sin{\left(2\pi Y\right)}\sin{\left(2\pi Z\right)}-\sin{\left(\pi\left(2X+2Y\right)\right)}\cos{\left(2\pi Z\right)}\right)\:.
\end{align}
However, the source term increases in complexity from Case I to Case
III, with Cases II and III consisting of on the order of $10^{4}$
and $10^{7}$ operations, respectively.

\begin{figure}
\begin{centering}
\includegraphics[width=0.95\textwidth,height=0.4\textheight,keepaspectratio]{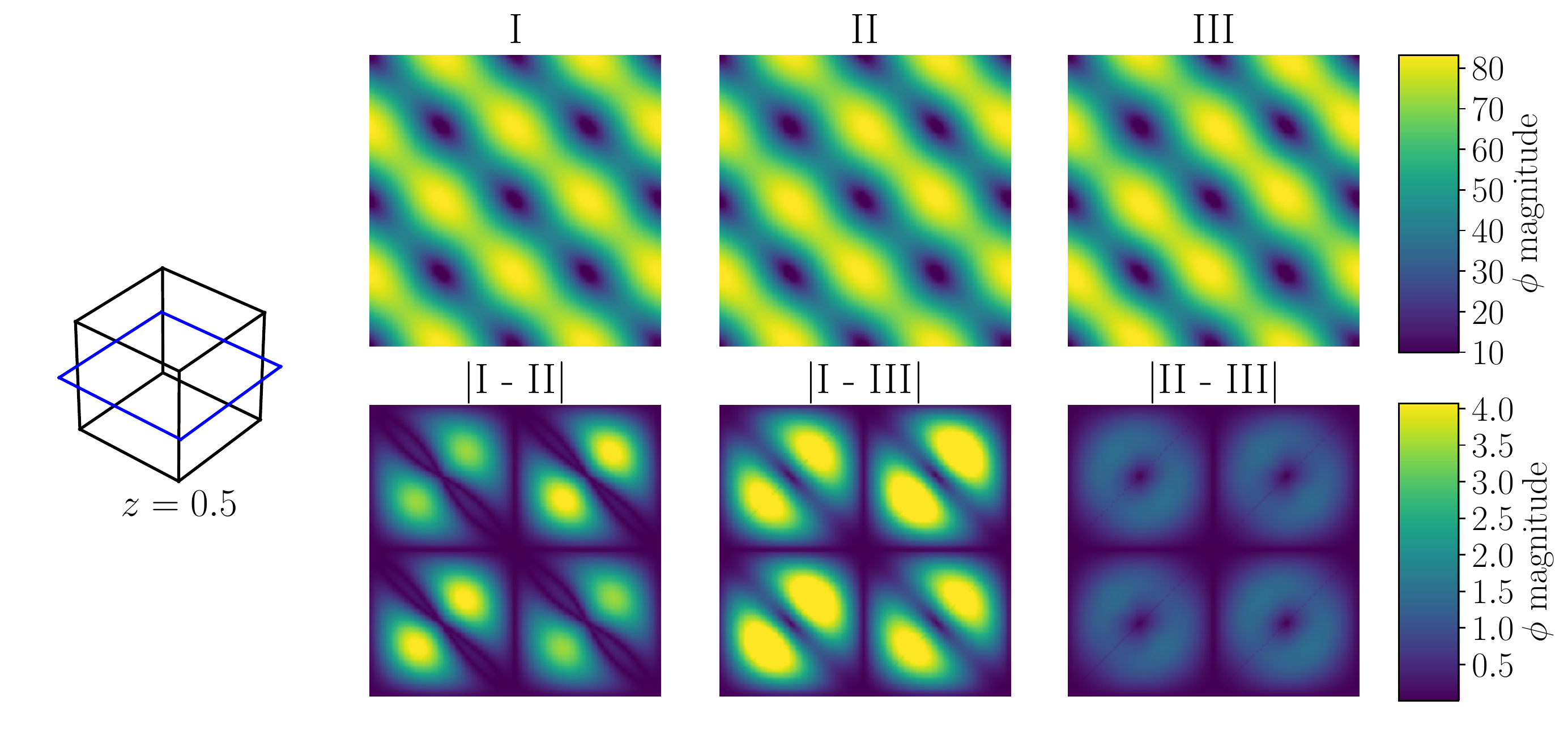}
\par\end{centering}
\begin{centering}
\includegraphics[width=0.95\textwidth,height=0.4\textheight,keepaspectratio]{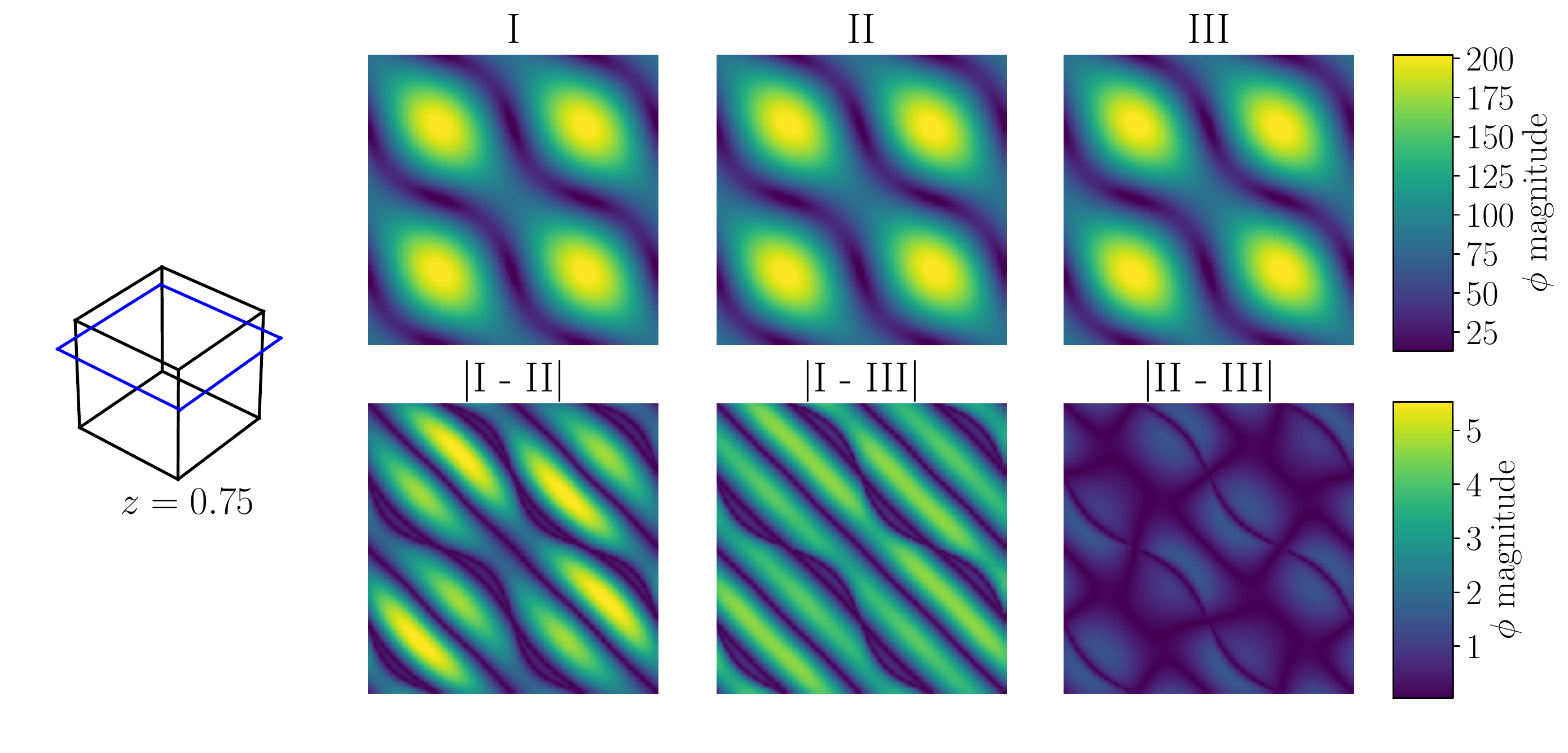}
\par\end{centering}
\caption{Contour plots of the source term magnitude on the $z=1/2$ and $z=3/4$
planes for Cases I–III. The magnitude difference between the source
terms for each case are also shown.\label{fig:source-term-fields} }
\end{figure}

Although the source term expressions differ greatly in mathematical
form, visualization of the source term fields shows that local differences
are relatively subtle (Fig. \ref{fig:source-term-fields}). Indeed,
maximal differences between Cases I and II and between Cases I and
III are only approximately 5\%, and differences between Cases II and
III are even less (approximately 1\%; Fig. \ref{fig:source-term-fields}).

\subsection{Error norms and observed orders of convergence}

\begin{figure}
\begin{centering}
\includegraphics[width=0.95\textwidth,height=0.4\textheight,keepaspectratio]{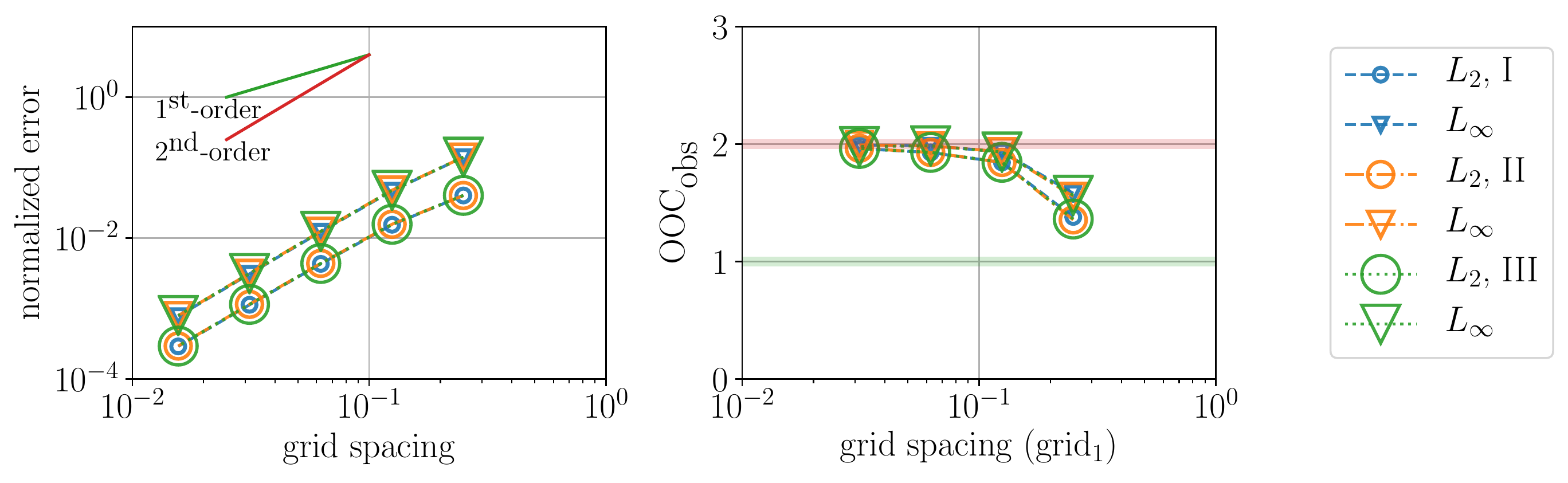}
\par\end{centering}
\caption{Normalized error norms (left) and observed orders of convergence ($\textrm{OOC}_{\textrm{obs}}$)
(right) for the three cases using \texttt{{*}DLOAD} and \texttt{C3D8I}
elements. Marker sizes are varied to allow for visualizing overlapping
data points. The $\textrm{OOC}_{\textrm{obs}}$ for all cases converges
towards two with decreasing grid spacing.\label{fig:OOC-DLOAD} Results
for \texttt{C3D8} and \texttt{C3D8R} element types are available online
as \protect\href{https://figshare.com/s/a67927162e674bbb791e}{Supplemental Material}.}
\end{figure}

\begin{figure}
\begin{centering}
\includegraphics[width=0.95\textwidth,height=0.4\textheight,keepaspectratio]{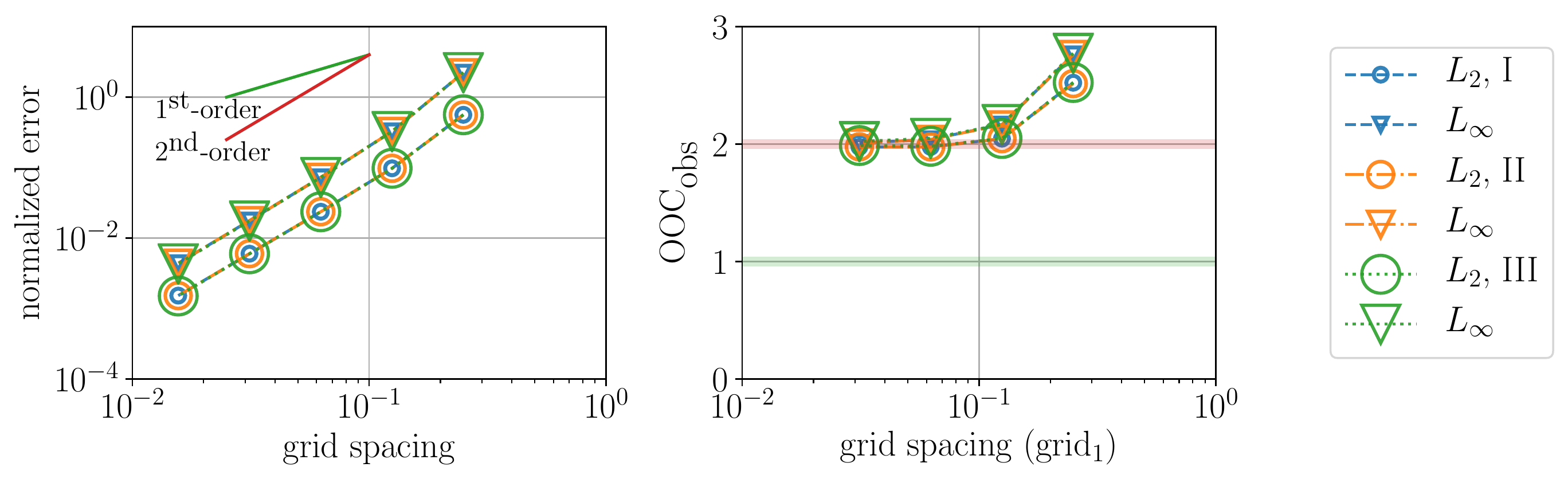}
\par\end{centering}
\caption{Normalized error norms (left) and observed orders of convergence ($\textrm{OOC}_{\textrm{obs}}$)
(right) the three cases using \texttt{{*}CLOAD} and \texttt{C3D8I}
elements. Marker sizes are varied to allow for visualizing overlapping
data points. The $\textrm{OOC}_{\textrm{obs}}$ for all cases converges
towards two with decreasing grid spacing.\label{fig:OOC-CLOAD} Results
for \texttt{C3D8} and \texttt{C3D8R} element types are available online
as \protect\href{https://figshare.com/s/a67927162e674bbb791e}{Supplemental Material}.}
\end{figure}

Spatial convergence behavior is examined qualitatively by plotting
error norms versus grid spacing on a $\log$-$\log$ scale and comparing
the slopes of the resulting lines to those representing example rates
of convergence (Figs. \ref{fig:OOC-DLOAD}–\ref{fig:OOC-CLOAD}, left).
For all cases investigated here, both the $L_{2}$ and $L_{\infty}$
error norms for displacement are reduced with mesh refinement at rates
that approach quadratic with increasing grid refinement (Figs. \ref{fig:OOC-DLOAD}–\ref{fig:OOC-CLOAD},
right; Tables \textbf{\ref{table:results_summary-grid-refinement-DLOAD}–\ref{table:results_summary-grid-refinement-CLOAD}}). 

Although both loading approaches yield similar convergence rates,
error norms are consistently an order of magnitude lower using\texttt{
{*}DLOAD} compared to \texttt{{*}CLOAD} (Fig. \ref{fig:OOC-DLOAD}
versus Fig. \ref{fig:OOC-CLOAD}). Convergence rates at coarse mesh
resolutions are lower than theoretical using \texttt{{*}DLOAD} but
higher than theoretical using \texttt{{*}CLOAD }(Table \textbf{\ref{table:results_summary-grid-refinement-DLOAD}}
versus Table\textbf{ \ref{table:results_summary-grid-refinement-CLOAD}}).
\begin{table}
\caption{Theoretical and observed orders of convergence ($\textrm{OOC}_{\textrm{theor}}$
and $\textrm{OOC}_{\textrm{obs}}$) in response to grid refinement
for the three MMS cases using \texttt{{*}DLOAD} and \texttt{C3D8I}
elements.\label{table:results_summary-grid-refinement-DLOAD} Results
for \texttt{C3D8} and \texttt{C3D8R} element types are available online
as \protect\href{https://figshare.com/s/a67927162e674bbb791e}{Supplemental Material}.}
\vspace{5mm}

\centering{}%
\begin{tabular}{cccccccc}
 &  & \multicolumn{6}{c}{$\textrm{OOC}_{\textrm{obs}}$}\tabularnewline
\cmidrule{3-8} 
$\textrm{grid}_{1}$ to $\textrm{grid}_{2}$  & $\textrm{OOC}_{\textrm{theor}}$ & $L_{2}$, I  & $L_{\infty}$, I  & $L_{2}$, II  & $L_{\infty}$, II  & $L_{2}$, III  & $L_{\infty}$, III \tabularnewline
\midrule
\midrule 
$4^{3}$ to $8^{3}$  & 2  & 1.38 & 1.58 & 1.36 & 1.55 & 1.36 & 1.57\tabularnewline
$8^{3}$ to $16^{3}$  & 2  & 1.84  & 1.93  & 1.84  & 1.94  & 1.85  & 1.93\tabularnewline
$16^{3}$ to $32^{3}$  & 2  & 1.93  & 1.99  & 1.93  & 1.99  & 1.93  & 1.99\tabularnewline
$32^{3}$ to $64^{3}$  & \textbf{2 } & \textbf{1.96 } & \textbf{1.99 } & \textbf{1.97 } & \textbf{1.99 } & \textbf{1.96} & \textbf{ 1.98}\tabularnewline
\end{tabular}
\end{table}
\begin{table}
\caption{Theoretical and observed orders of convergence ($\textrm{OOC}_{\textrm{theor}}$
and $\textrm{OOC}_{\textrm{obs}}$) in response to grid refinement
for the three MMS cases using \texttt{{*}CLOAD} and \texttt{C3D8I}
elements.\label{table:results_summary-grid-refinement-CLOAD} Results
for \texttt{C3D8} and \texttt{C3D8R} element types are available online
as \protect\href{https://figshare.com/s/a67927162e674bbb791e}{Supplemental Material}.}
\vspace{5mm}

\centering{}%
\begin{tabular}{cccccccc}
 &  & \multicolumn{6}{c}{$\textrm{OOC}_{\textrm{obs}}$}\tabularnewline
\cmidrule{3-8} 
$\textrm{grid}_{1}$ to $\textrm{grid}_{2}$  & $\textrm{OOC}_{\textrm{theor}}$ & $L_{2}$, I  & $L_{\infty}$, I  & $L_{2}$, II  & $L_{\infty}$, II  & $L_{2}$, III  & $L_{\infty}$, III \tabularnewline
\midrule
\midrule 
$4^{3}$ to $8^{3}$  & 2  & 2.52  & 2.76  & 2.52  & 2.76  & 2.53  & 2.77 \tabularnewline
$8^{3}$ to $16^{3}$  & 2  & 2.05  & 2.17  & 2.05  & 2.16  & 2.05  & 2.17 \tabularnewline
$16^{3}$ to $32^{3}$  & 2  & 1.98  & 2.04  & 1.98  & 2.04  & 1.98  & 2.05 \tabularnewline
$32^{3}$ to $64^{3}$  & \textbf{2 } & \textbf{1.98 } & \textbf{2.01 } & \textbf{1.98 } & \textbf{2.01 } & \textbf{1.99 } & \textbf{2.02 }\tabularnewline
\end{tabular}
\end{table}

The pseudo-time increment refinement study confirms that the increment
size has no influence on the results when using the small-strain formulation
(Case I) and a relatively small but quantifiable influence on the
results when using the finite-strain formulation (Cases II and III).
For the finite strain cases, the observed $p$-value for increment
size is remarkably consistent and close to unity ($1.02\pm0.01$),
even for the coarsest increment size triplet (Table \textbf{\ref{table:results_summary-increment-refinement}}).
\begin{table}
\caption{Observed convergence rate ($p$) in response to increment refinement
for the two finite-strain, updated Lagrangian MMS cases using \texttt{{*}CLOAD}
and \texttt{C3D8I} elements.\label{table:results_summary-increment-refinement}}

\centering{}\vspace{5mm}
\begin{tabular}{ccccc}
\cmidrule{2-5} 
 & \multicolumn{4}{c}{observed $p$}\tabularnewline
\cmidrule{2-5} 
increment size triplet & $L_{2}$, II  & $L_{\infty}$, II  & $L_{2}$, III  & $L_{\infty}$, III \tabularnewline
\midrule
\midrule 
0.2–0.1–0.05 & 1.02 & 1.02 & 1.03 & 1.03\tabularnewline
0.1–0.05–0.025 & 1.04 & 1.04 & 1.01 & 1.01\tabularnewline
0.05–0.025–0.0125 & 1.03 & 1.03 & 1.01 & 1.01\tabularnewline
0.025–0.0125–0.00625 & 1.02 & 1.01 & 1.00 & 1.01\tabularnewline
0.0125–0.00625–0.003125 & 1.01 &  1.01 &  1.00 &  1.00\tabularnewline
\end{tabular}
\end{table}

\section{Discussion and Conclusions}

In this study, we perform MMS code verification of elastostatic solid
mechanics problems for three distinct cases: I) small-strain linear-elasticity,
II) finite-strain neo-Hookean hyperelasticity, and III) finite-strain
Hencky elasticity. These cases use simple but common constitutive
models that have relevance to many applications, including cardiovascular
and orthopedic medical devices. In particular, most plastic and pseudoplastic
or superelastic models are built upon finite-strain Hencky elasticity.

Observed orders of convergence in response to grid refinement approach
quadratic for all cases (Figs. \ref{fig:OOC-DLOAD}–\ref{fig:OOC-CLOAD},
right; Tables \textbf{\ref{table:results_summary-grid-refinement-DLOAD}}–\textbf{\ref{table:results_summary-grid-refinement-CLOAD}}),
agreeing closely with theory for displacement-based linear finite
elements \cite{bathe2006finite,curnier2012computational}. The observed
convergence rates in response to increment refinement for the finite-strain
cases are also remarkably consistent and close to unity (Table \textbf{\ref{table:results_summary-increment-refinement}}),
providing evidence of proper implementation of the updated Lagrangian
method in \texttt{ABAQUS}. Collectively, these results provide evidence
that coding errors do not negatively influence the simulation results
for these particular classes of elastostatic problems using version
R2016x of \texttt{ABAQUS} in our particular computing environment
(a Microsoft Windows 7 workstation).

\begin{figure}
\begin{centering}
\includegraphics[width=0.95\textwidth,height=0.4\textheight,keepaspectratio]{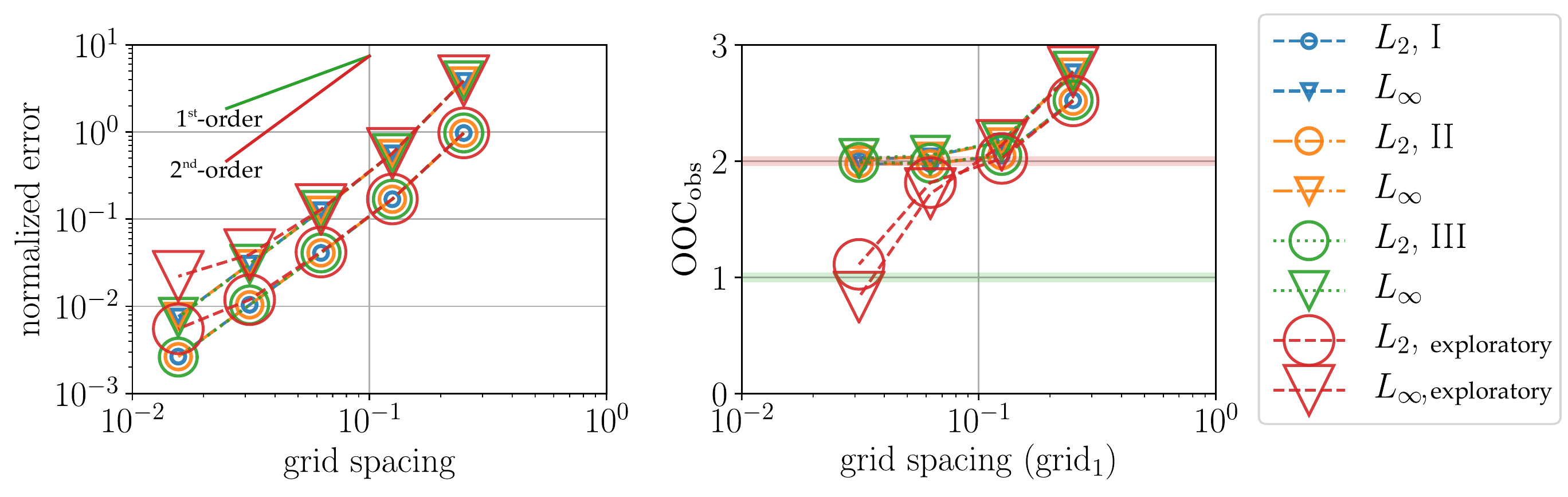}
\par\end{centering}
\caption{Normalized error norms (left) and observed orders of convergence ($\textrm{OOC}_{\textrm{obs}}$)
(right) for the three cases and the exploratory case using \texttt{{*}CLOAD}
and \texttt{C3D8I} elements. Marker sizes are varied to allow for
visualizing overlapping data points. The exploratory case was performed
using the finite-strain Hencky-elastic constitutive model but applying
the neo-Hookean source terms. The exploratory case demonstrates the
sensitivity of MMS to errors, as the maximum differences between the
source terms for Cases II and III are on the order of only $1\%$.\label{fig:OOC-check-case}}
\end{figure}

Interestingly, although the analytical source terms vary by orders
of magnitude in the number of mathematical operations they contain,
qualitatively the source terms are quite similar (Fig. \ref{fig:source-term-fields}).
To demonstrate the sensitivity of MMS to these subtle differences,
an additional exploratory case is performed pairing the constitutive
model from Case III (finite-strain Hencky-elasticity) with the source
terms from Case II (neo-Hookean hyperelasticity). Although the source
terms for these two cases differ at most by only $\approx1\%$ (e.g.,
Fig. \ref{fig:source-term-fields}), the erroneous pairing of the
source term and constitutive model causes the convergence rate to
fall off sharply after the grid spacing decreases below $10^{-1}$
(Fig. \ref{fig:OOC-check-case}; Table \ref{table:results_summary-grid-refinement-check-case}).
Accordingly, this illustrates the sensitivity of the MMS technique
to subtle errors in the implementation of constitutive models, boundary
conditions, body forces, or any other alteration to the underlying
mathematical model, making the method a powerful tool for code testing
and development. 
\begin{table}
\caption{Theoretical and observed orders of convergence ($\textrm{OOC}_{\textrm{theor}}$
and $\textrm{OOC}_{\textrm{obs}}$) in response to grid refinement
for the exploratory case (Case III constitutive model with Case II
source term) using \texttt{{*}CLOAD} and \texttt{C3D8I} elements.
Note that the $\textrm{OOC}_{\textrm{obs}}$ does not converge to
the theoretical value of two despite the subtle quantitative differences
between the Case II and Case III source terms.\label{table:results_summary-grid-refinement-check-case}}

\centering{}\vspace{5mm}
\begin{tabular}{cccc}
\cmidrule{3-4} 
 &  & \multicolumn{2}{c}{$\textrm{OOC}_{\textrm{obs}}$}\tabularnewline
\cmidrule{3-4} 
$\textrm{grid}_{1}$ to $\textrm{grid}_{2}$  & $\textrm{OOC}_{\textrm{theor}}$ & $L_{2}$  & $L_{\infty}$\tabularnewline
\midrule
\midrule 
$4^{3}$ to $8^{3}$  & 2  & 2.52  & 2.77 \tabularnewline
$8^{3}$ to $16^{3}$  & 2  & 2.02  & 2.14 \tabularnewline
$16^{3}$ to $32^{3}$  & 2  & 1.82  & 1.72 \tabularnewline
$32^{3}$ to $64^{3}$  & \textbf{2 } & \textbf{1.11 } & \textbf{0.83 }\tabularnewline
\end{tabular}
\end{table}

We note that the use of commercial software poses some unique challenges
for users desiring to perform code verification \cite{salari2000code}.
First, although code verification is often described as gathering
evidence that the ``equations are solved correctly'' \cite{roache2009fundamentals},
with commercial software one must first answer the question: ``what
equations are being solved?'' Answering this question is not always
trivial. For example, initial attempts at MMS here for the simple
small-strain linear-elastic problem failed (i.e., $\textrm{OOC}_{\textrm{obs}}$
tending towards zero with increasing mesh refinement) due to the use
of the wrong governing equations in the generation of the analytical
source term. Closer inspection of the documentation revealed that
\texttt{ABAQUS} makes an additional approximation to the governing
equations when simulating infinitesimal strain problems (Eqn. \ref{eq:small-strain-approximation};
\cite{simulia2016abaqus}). After accounting for this approximation,
the results were greatly improved and second-order displacement convergence
rates were observed (Table \textbf{\ref{table:results_summary-grid-refinement-CLOAD}}).
The neo-Hookean case was also initially challenging due to the unique
form of the constitutive model used by \texttt{ABAQUS} \cite{simulia2016abaqus}.
Thus, determining the correct underlying governing equations for a
commercial software package requires a user to at least carefully
and thoroughly study the documentation. When questions arise that
are not answered by the documentation alone, communication with the
vendor is required to gain clarification on the precise form of the
governing equations implemented in the software. Otherwise, as demonstrated
here with our exploratory case, MMS code verification of the software
will likely fail. 

Second, the addition of an analytical source term into the governing
equations is not always straightforward when using commercial software
\cite{salari2000code,roache2002code,roache2009fundamentals}. For
example, during our initial attempts at using \texttt{{*}DLOAD} for
MMS, we encountered relatively low initial error norms at the coarsest
mesh level but observed convergence rates that approached zero (rather
than two) with increasing mesh refinement. By running numerical experiments
we discovered that the behavior of \texttt{{*}DLOAD }differs under
the small- and finite-strain formulations (see Methods, Eqn. \ref{eq:dload-and-nlgeom}),
and correct convergence rates were observed after accounting for this
detail. For simplicity, we also used concentrated point loads (\texttt{{*}CLOAD)}
weighted by their respective nodal volumes as an alternative to the
built-in distributed load subroutine. In doing so, we were able to
implement the analytical source term in a transparent manner. Similar
strategies may also prove helpful in other situations wherein users
cannot easily add a source term to the governing equations. Salari
and Knupp also provide additional suggestions for implementing source
terms when working with commercial software (\cite{salari2000code},
Appendix B). 

Finally, although code verification activities provide valuable information
to code developers, user-led code verification activities for commercial
software has limited utility if problems are discovered. For instance,
if a user finds an order of accuracy error in a commercial software,
the user cannot investigate the source code to diagnose the problem
or to make improvements—they can only submit a bug report to the software
vendor in hopes that the vendor will investigate and correct the issue
in a future software release. Given the limitations and difficulties
associated with user-led code verification, one may argue that software
vendors should bear the primary responsibility for performing rigorous
code verification. Unfortunately, in practice, vendors cannot anticipate
and verify every possible combination of software options that users
may invoke. Many simulation packages also allow users to write their
own coding extensions for custom boundary conditions or constitutive
models (e.g., \texttt{ABAQUS UMAT}), and it would be unreasonable
to expect vendors to take responsibility for verifying user-written
code customizations. Indeed, code verification activities are inherently
application-specific. Nevertheless, Oberkampf and Roy \cite{oberkampf2010verification}
state that although the end-user holds the ultimate responsibility
for code verification, ``commercial software companies are unlikely
to perform rigorous code verification studies unless users request
it.'' Accordingly, advancing code verification practices for commercial
software will likely require close collaborations between users and
software developers. Releasing the documentation and source code for
MMS studies as open-source software, as we do here, will allow other
users to extend MMS for their applications and will facilitate the
wider adoption of this powerful approach for code verification.

In conclusion, we provide rigorous code verification evidence for
elastostatic analyses with three common constitutive equations using
a commercial finite element solver (\texttt{ABAQUS}). As mentioned
above, code verification exercises can only provide verification evidence
for a single set of governing equations, and changing the form of
a single term in the governing equations triggers the need to gather
additional verification evidence \cite{roache2002code,roache2009fundamentals}.
Hence, additional verification evidence should be generated when performing
simulations with different constitutive models or when considering
additional physics such as contact. With this in mind, we have provided
the code used herein for generating analytical source terms as supplemental
material in the form of a \texttt{Python Jupyter} notebook, available
online at \href{https://figshare.com/s/a67927162e674bbb791e}{https://figshare.com/s/a67927162e674bbb791e},
in the hope that it will serve as a starting template for others who
desire to perform MMS code verification for their own specific applications.\textcolor{red}{}

\section*{Conflict of interest statement}

One of the authors (NR) was formerly an employee of Dassault Systèmes
Simulia, makers of \texttt{ABAQUS}.

\section*{Acknowledgments}

\noindent We thank Robert L. Campbell (Pennsylvania State University)
and Scott T. Miller (Sandia National Laboratory) for helpful discussions.
Thanks also to Joshua E. Soneson and Tina M. Morrison (U.S. FDA) for
reviewing the manuscript. This study was funded by the U.S. FDA Center
for Devices and Radiological Health (CDRH) Critical Path program.
The research was supported in part by an appointment to the Research
Participation Program at the U.S. FDA administered by the Oak Ridge
Institute for Science and Education through an interagency agreement
between the U.S. Department of Energy and FDA. The findings and conclusions
in this article have not been formally disseminated by the U.S. FDA
and should not be construed to represent any agency determination
or policy. The mention of commercial products, their sources, or their
use in connection with material reported herein is not to be construed
as either an actual or implied endorsement of such products by the
Department of Health and Human Services.\clearpage{}

\bibliographystyle{/Users/kenneth.aycock/Desktop/Manuscripts/2017_CP_01_3D_Printing_PIV_manuscript/OLD/JBME_Template_journals-1.0/asmems4}
\bibliography{Aycock_Rebelo_Craven_Code_Verification_for_Elastostatics_Problems}

\begin{thebibliography}{10}

\bibitem{morrison2018advancing}
Morrison, T.~M., Pathmanathan, P., Adwan, M., and Margerrison, E., 2018.
\newblock ``Advancing regulatory science with computational modeling for
  medical devices at the {FDA}'s {Office of Science and Engineering
  Laboratories}''.
\newblock {\em Frontiers in Medicine, {\bf 5}}.

\bibitem{himes2016augmenting}
Himes, A., Haddad, T., and Bardot, D., 2016.
\newblock ``Augmenting a clinical study with virtual patient models: {Food and
  Drug Administration} and industry collaboration''.
\newblock {\em Journal of Medical Devices, {\bf 10}}(3), p.~030947.

\bibitem{himes2018use}
Himes, A., 2018.
\newblock ``The use of computational modeling and simulation to create virtual
  patients: Application to cardiac pacing and defibrillation systems''.
\newblock {\em Journal of Cardiovascular Translational Research, {\bf 11}}(2),
  April, pp.~89--91.

\bibitem{haddad2014fracture}
Haddad, T., Himes, A., and Campbell, M., 2014.
\newblock ``Fracture prediction of cardiac lead medical devices using
  {Bayesian} networks''.
\newblock {\em Reliability Engineering \& System Safety, {\bf 123}},
  pp.~145--157.

\bibitem{taylor2013computational}
Taylor, C.~A., Fonte, T.~A., and Min, J.~K., 2013.
\newblock ``Computational fluid dynamics applied to cardiac computed tomography
  for noninvasive quantification of fractional flow reserve: Scientific
  basis''.
\newblock {\em Journal of the American College of Cardiology, {\bf 61}}(22),
  pp.~2233--2241.

\bibitem{zarins2013computed}
Zarins, C.~K., Taylor, C.~A., and Min, J.~K., 2013.
\newblock ``Computed fractional flow reserve ({$\textrm{FFT}_\textrm{CT}$})
  derived from coronary {CT} angiography''.
\newblock {\em Journal of Cardiovascular Translational Research, {\bf 6}}(5),
  pp.~708--714.

\bibitem{ASME_VV40}
{{ASME V\&V} 40 -- 2018}.
\newblock {\em Assessing credibility of computational models through
  verification and validation: application to medical devices}.
\newblock {American Society of Mechanical Engineers}.

\bibitem{ASME-VV10}
{{ASME V\&V 10} -- 2006 ({R}2016)}.
\newblock {\em Guide for verification and validation in computational solid
  mechanics}.
\newblock {American Society of Mechanical Engineers}.

\bibitem{ASME_VV20}
{{ASME V\&V 20} -- 2009 ({R}2016)}.
\newblock {\em Standard for verification and validation in computational fluid
  dynamics and heat transfer}.
\newblock {American Society of Mechanical Engineers}.

\bibitem{oberkampf2010verification}
Oberkampf, W.~L., and Roy, C.~J., 2010.
\newblock {\em Verification and validation in scientific computing}.
\newblock Cambridge University Press.

\bibitem{roache2009fundamentals}
Roache, P.~J., 2009.
\newblock {\em Fundamentals of verification and validation}.
\newblock Hermosa Albuquerque, NM.

\bibitem{roache2002code}
Roache, P.~J., 2002.
\newblock ``Code verification by the method of manufactured solutions''.
\newblock {\em Journal of Fluids Engineering, {\bf 124}}(1), pp.~4--10.

\bibitem{roache1972computational}
Roache, P.~J., 1972.
\newblock {\em Computational fluid dynamics}.
\newblock Hermosa publishers.

\bibitem{steinberg1985symbolic}
Steinberg, S., and Roache, P.~J., 1985.
\newblock ``Symbolic manipulation and computational fluid dynamics''.
\newblock {\em Journal of Computational Physics, {\bf 57}}(2), pp.~251--284.

\bibitem{oberkampf1995methodology}
Oberkampf, W., Blottner, F., and Aeschliman, D., 1995.
\newblock ``Methodology for computational fluid dynamics code
  verification/validation''.
\newblock In Fluid Dynamics Conference, p.~2226.

\bibitem{bathe1990displacement}
Bathe, K.-J., Luiz~Bucalem, M., and Brezzi, F., 1990.
\newblock ``Displacement and stress convergence of our {MITC} plate bending
  elements''.
\newblock {\em Engineering Computations, {\bf 7}}(4), pp.~291--302.

\bibitem{batra1997finite}
Batra, R., and Liang, X., 1997.
\newblock ``Finite dynamic deformations of smart structures''.
\newblock {\em Computational Mechanics, {\bf 20}}(5), pp.~427--438.

\bibitem{batra2006consideration}
Batra, R., and Love, B., 2006.
\newblock ``Consideration of microstructural effects in the analysis of
  adiabatic shear bands in a tungsten heavy alloy''.
\newblock {\em International Journal of Plasticity, {\bf 22}}(10),
  pp.~1858--1878.

\bibitem{Chamberland_2010}
Chamberland, {\'E}., Fortin, A., and Fortin, M., 2010.
\newblock ``Comparison of the performance of some finite element
  discretizations for large deformation elasticity problems''.
\newblock {\em Computers \& Structures, {\bf 88}}(11-12), Jun, pp.~664--673.

\bibitem{kamojjala2015verification}
Kamojjala, K., Brannon, R., Sadeghirad, A., and Guilkey, J., 2015.
\newblock ``Verification tests in solid mechanics''.
\newblock {\em Engineering with Computers, {\bf 31}}(2), pp.~193--213.

\bibitem{salari2000code}
Salari, K., and Knupp, P., 2000.
\newblock Code verification by the method of manufactured solutions.
\newblock Tech. rep., Sandia National Labs., Albuquerque, NM (US); Sandia
  National Labs., Livermore, CA (US).

\bibitem{xiao2002hencky}
Xiao, H., and Chen, L., 2002.
\newblock ``Hencky's elasticity model and linear stress-strain relations in
  isotropic finite hyperelasticity''.
\newblock {\em Acta Mechanica, {\bf 157}}(1-4), pp.~51--60.

\bibitem{xiao2005hencky}
Xiao, H., 2005.
\newblock ``Hencky strain and {Hencky} model: extending history and ongoing
  tradition''.
\newblock {\em Multidiscipline Modeling in Materials and Structures, {\bf
  1}}(1), pp.~1--52.

\bibitem{Truesdell1960}
Truesdell, C., and Toupin, R., 1960.
\newblock {\em The Classical Field Theories}.
\newblock Springer Berlin Heidelberg, Berlin, Heidelberg, pp.~226--858.

\bibitem{bathe2006finite}
Bathe, K.-J., 2006.
\newblock {\em Finite element procedures}.
\newblock Klaus-Jurgen Bathe.

\bibitem{simulia2016abaqus}
{Simulia, Dassault Systemes}, 2016.
\newblock ``Abaqus 2016x documentation''.
\newblock {\em Providence, Rhode Island, US}.

\bibitem{elsworth2014verification}
Elsworth, C.~W., 2014.
\newblock ``Verification of an overset-grid enabled fluid-structure interaction
  solver''.
\newblock Master's thesis, The Pennsylvania State University.

\bibitem{gurtin1982introduction}
Gurtin, M.~E., 1982.
\newblock {\em An introduction to continuum mechanics}, Vol.~158.
\newblock Academic press.

\bibitem{wolfram2011mathematica}
{Wolfram}, 2011.
\newblock ``Mathematica 11 documentation''.
\newblock {\em Wolfram Research Inc., March}.

\bibitem{bavzant2012work}
Ba{\v{z}}ant, Z.~P., Gattu, M., and Vorel, J., 2012.
\newblock ``Work conjugacy error in commercial finite-element codes: its
  magnitude and how to compensate for it''.
\newblock {\em Proceedings of the Royal Society-A, {\bf 468}}(2146), p.~3047.

\bibitem{curnier2012computational}
Curnier, A., 2012.
\newblock {\em Computational methods in solid mechanics}, Vol.~29.
\newblock Springer Science \& Business Media.

\end{thebibliography}

\end{document}